\documentclass[sigconf]{acmart}
\usepackage{booktabs} % For formal tables

\usepackage{xspace}

\newcommand{\system}{\textsc{Graph Playground}\xspace}
\newcommand{\overview}{Overview\xspace}
\newcommand{\ribbon}{Ribbon\xspace}
\newcommand{\layers}{Layers\xspace}
\newcommand{\user}{Don\xspace}
\newcommand{\spnet}{shortest-path-net\xspace}
\usepackage{makecell}
\usepackage{enumitem}
\usepackage{tablefootnote}
\usepackage{quoting}

\setcopyright{none}
\renewcommand\footnotetextcopyrightpermission[1]{} % removes footnote with conference information in first column
\begin{document}
\title{Large Graph Exploration via\\Subgraph Discovery and Decomposition}

\author{James Abello}
\authornote{Authors contributed equally.}
\affiliation{%
  \institution{Rutgers University}
  \city{New Brunswick}
  \state{New Jersey}
}
\email{abelloj@cs.rutgers.edu}

\author{Fred Hohman}
\authornotemark[1]
\affiliation{%
  \institution{Georgia Institute of Technology}
  \city{Atlanta}
  \state{Georgia}
}
\email{fredhohman@gatech.edu}

\author{Varun Bezzam}
\affiliation{%
  \institution{Georgia Institute of Technology}
  \city{Atlanta}
  \state{Georgia}
}
\email{varun.bezzam@gatech.edu}

\author{Duen Horng Chau}
\affiliation{%
  \institution{Georgia Institute of Technology}
  \city{Atlanta}
  \state{Georgia}
}
\email{polo@gatech.edu}

% The default list of authors is too long for headers.
% \renewcommand{\shortauthors}{B. Trovato et al.}

\begin{abstract}
We are developing an interactive graph exploration system called \system for making sense of large graphs. 
\system offers a fast and scalable edge decomposition algorithm, based on iterative vertex-edge peeling, to decompose million-edge graphs in seconds.
\system introduces a novel graph exploration approach and a 3D representation framework that  simultaneously reveals (1) peculiar subgraph structure discovered through the decomposition's layers, (e.g., quasi-cliques), and (2) possible vertex roles in linking such subgraph patterns across layers.
\end{abstract}


\maketitle

\section{Introduction}
\label{sec:intro}

Graphs are everywhere, growing increasingly complex, and still lack scalable, interactive tools to support sensemaking. 
In a recent online survey conducted to gather information about how graphs are used in practice, 
graph analysts rated scalability and visualization as the most pressing issues to address~\cite{sahu2017ubiquity}.
While graph drawing techniques have been developed to improve the layout of a graph in 2D, 
these approaches become less effective when visualizing modern day large graphs.
As a response, advanced approaches such as ``super-noding''~\cite{abello2006ask, archambault2007grouse, archambault2007topolayout}, and edge bundling~\cite{holten2006hierarchical, cui2008geometry, alper2013weighted} have been designed to visually reduce the number of glyphs visible to a user.
Some work abstracts graphs to higher-level representations, such as using contours and heat maps as a proxy for vertex density~\cite{lin2010contextour, cao2010facetatlas}, graph motifs for repeating structural patterns~\cite{dunne2013motif}, and overall graph summarizations~\cite{koutra2014vog}.
New modes of exploration based on relevance and measures of ``interestingness'' have also been developed to explore large graphs without showing every vertex and edge~\cite{pienta2017facets, chau2011apolo, heer2004doitrees}.
While these approaches may help users develop insights into a graph's functional properties, scalability, interaction, and extracting overall descriptive information about an unknown graph as it is being explored remain pressing issues 
in large graph exploration systems. 

\textit{Edge decomposition algorithms}, based on fixed points of degree peeling, show strong potential for helping users explore unfamiliar graph data~\cite{hohman2017playground, abello2014network}, because  
(1) they can discover peculiar subgraph patterns structurally similar or dissimilar to regular subgraphs;
(2) they can quantify possible ``roles'' a vertex can play in the overall network topology;
and (3) they scale to large graphs.

In this ongoing work,
we show how using scalable edge decompositions~\cite{abello2014network} as a central
mechanism for navigation, exploration, and large data sensemaking can reveal interesting graph structure previously unknown to users.
Our fast and scalable edge decomposition divides large graphs into an ordered set of \textit{graph layers}
that is dependent only upon the topology of the graph.
In this decomposition, edges are unique and participate in particular layers; however, vertices can be duplicated and exist in multiple layers at once; we call these vertices \textit{clones}.
Graph layers help users identify potentially important substructures (e.g., quasi-cliques, multi-partite-cores), 
by automatically separating such patterns from the majority of the graph, while vertex clones allow one to link related layers together using \textit{cross-layer exploration}.
Together, we introduce \system (\autoref{fig:ui}), an interactive graph exploration system that 
decompose large graphs quickly, generating explorable multi-layered representations that help graph data analysts interactively discover and make sense of peculiar graph structures.
Through \system, we contribute:

\begin{itemize}[leftmargin=3mm,topsep=3pt]
    \setlength\itemsep{0em}
    
    \item \textbf{New paradigm for graph exploration and navigation.}
    We propose a new paradigm for graph exploration and navigation centered around two novel components produced by our edge decomposition algorithm: graph layers and vertex clones.
    Graph layers are topologically and structurally interesting subgraphs of the original graph that can be analyzed independently; however, using vertex clones, vertices that exist in multiple layers, allows one to explore a graph across layers by providing a means to navigate local structure with a global context.
    
    \item \textbf{Fast, scalable edge decomposition via memory mapping and multithreading.}
    We present a fast and scalable edge decomposition algorithm using memory mapped I/O and multithreaded processes.
    We present decompositions on a wide range of graphs, varying in both size (e.g., up to hundreds of millions of edges) and domain (e.g., social networks, hyperlink networks, and co-occurrence networks) and tabulate computational timings and structural results.
    We can decompose graphs with millions of edges in seconds, and graphs with hundreds of millions of edges in minutes.
    
    \item \textbf{\system.}
    \system is a web-based interactive graph visualization system composed of three main linked views to help users explore and navigate large graphs.
    It uses GPUs for large force-directed graph layouts as well as accelerated 3D graphics to demonstrate how the edge decomposition divides the original graph into layers.
    \system simultaneously reveals ``peculiar'' subgraph structure discovered through the decomposition's layers, (e.g., quasi-cliques), and ``possible'' vertex roles in linking such subgraph patterns across layers.

\end{itemize}

\section{Illustrative Scenario}
\label{sec:scenario}

To illustrate how \system can help users explore large graphs and discover interesting structure, consider our user \user who wants to explore and make sense of a word embedding graph generated from Wikipedia from 2014.
Word embeddings are an increasingly popular and important technique that turns words into high dimensional vectors (e.g., 300 dimensions)~\cite{bengio2003neural, mikolov2013efficient, pennington2014glove}.
These word embeddings are used as input to many machine learning applications such as visual question answering~\cite{antol2015vqa} and neural machine translation~\cite{bahdanau2014neural}.
Therefore it is important to make sense of what information a word embedding has captured and how well the embedding matches our understanding of language.

\user's Wikipedia word embedding graph is generated using word vectors from GloVe, an unsupervised learning algorithm for obtaining vector representations for words~\cite{pennington2014glove}.
The graph contains 65,870 vertices and 213,526 edges.
Each vertex is a unique word, 
and an edge connects two words if the angular distance between their two word vectors is less than some threshold.\footnote{Angular distance is closely related to cosine similarity, and is an effective method for measuring the linguistic or semantic similarity of corresponding words~\cite{pennington2014glove}.
For this graph, the threshold to connect two words is set to 0.9.
Words with numbers/digits are removed from the dataset and are not considered.
}

\begin{figure*}[t]
 \centering
 \includegraphics[width=\textwidth]{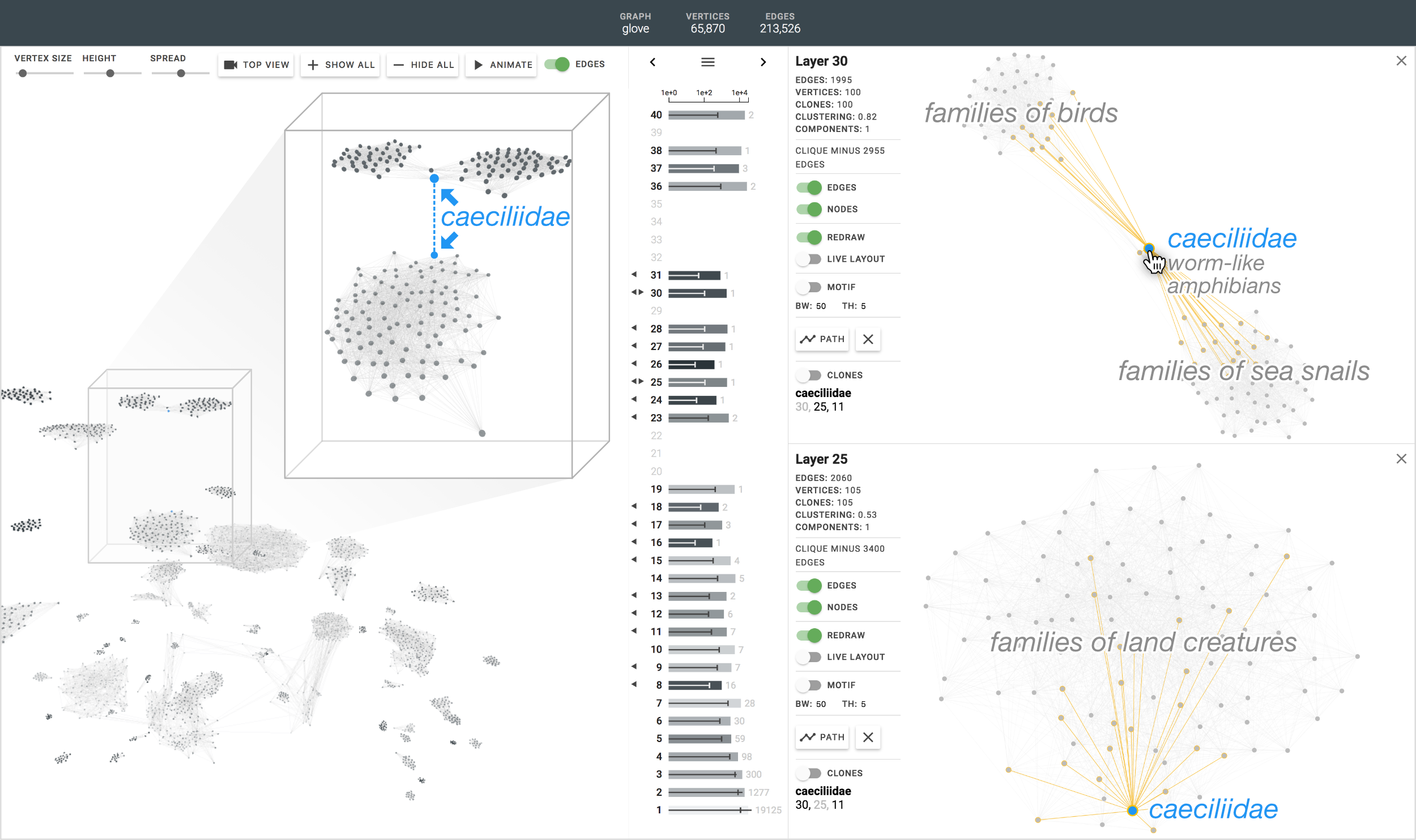}
 \caption{
 The \system user interface.
 \system is composed of three main views: the 3D \overview (left), the \ribbon (middle), and the \layers view (right).
 The \ribbon that splits the display can be dragged left and right to adjust the visible screen real estate that either the \overview or \layers view shows.
 In the figure, the vertex ``caeciliidae'' is selected, coloring it blue in both the \overview and \layers view.
 Here we see ``caeciliidae'' (a worm-like amphibian) in layer 30 bridges two quasi-cliques (families of birds and families of sea snails) together, while its clone in layer 25 participates in another single quasi-clique (families of land creatures).}
 \label{fig:ui}
\end{figure*}

\textbf{Visualizing edge decompositions.}
\user is exploring the word embedding graph for the first time; therefore, he first wants to see a high-level, global representation of the graph.
The \overview (\autoref{fig:ui}, left) is one of three main views in the \system user interface and visualizes a natural 3D representation of the edge decomposition's output that assigns each found graph layer a height based on its layer value.
\user adjusts the vertical separation between layers, to better visualize patterns revealed by the decomposition.
In the \overview, denser layers rise to the top of the 3D structure (e.g., quasi-cliques), while spare structures sink to lower layers (e.g., trees, stars). 

\textbf{Finding interesting graph layers.}
\user now wants a more quantitative view of the graph.
He inspects the graph \ribbon
(\autoref{fig:ui}, middle), where each graph layer is encoded by a glyph that visualizes the graph layer's edge count, vertex count, clone count, the number of connected components, and the clustering coefficient.
The \ribbon provides a compact, information-rich summarization of the edge decomposition using well-studied graph measures.
\user can now more clearly see how many layers this graph has (31 total layers, with the highest value being 40), and how certain measures, such as their clustering coefficient density (bar color), vary over the layers.

\begin{figure}[t]
 \centering
 \includegraphics[width=1.0\linewidth]{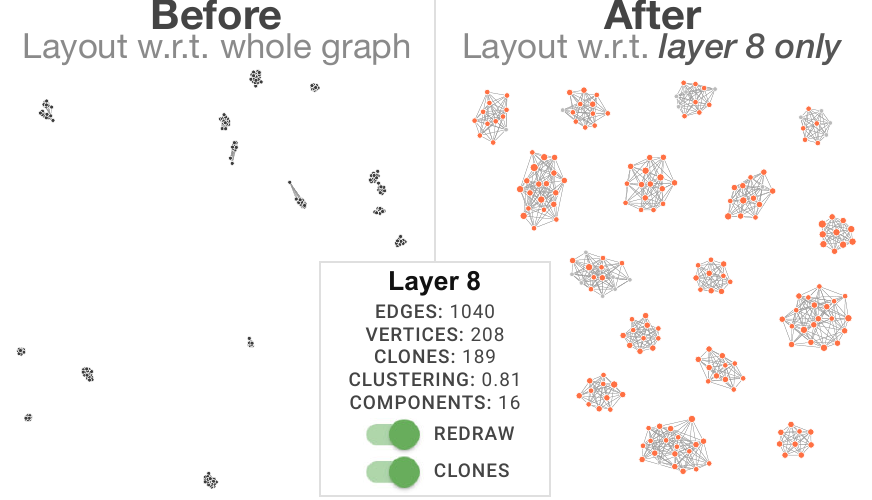}
 \caption{
 Layer 8 from the word embedding graph.
 On the left is the original layout computed with respect to the entire graph, but now that we have separated out layer 8 from the remaining graph, we can recompute its layout independently.
 This produces the layout on the right, where the cloned vertices are colored red and sized according to how many clones they have in the remainder of the graph.
 }
 \label{fig:scenario}
\end{figure}

\user finds layer 8 interesting, because it is a highly dense layer, but it is further down in the \ribbon than the other dense layers.
\user clicks the 8th graph layer glyph in the \ribbon.
\system now displays layer 8 in the 2D \layers view (\autoref{fig:ui}, right).
\user is presented with a handful of small tangled connected components (\autoref{fig:scenario}, left); however, the layouts of these components were computed with respect to the entire graph, but since we are only visualizing a particular layer from the edge decomposition, \system enables \user to perform a force-directed layout with respect to only this layer.
When this is performed, \user watches as all the small components animate and reveal that layer 8 is a collection of highly dense small quasi-cliques (\autoref{fig:scenario}, right).
This view is fully interactive: \user can zoom and pan over the graph layer, hovering over a vertex displays the vertex's label and highlights its immediate neighbors, and vertices can be selected and dragged around for maximum control over the graph layout.

\textbf{Cross-layer exploration.}
Recall that \system discovered a small collection of quasi-cliques in layer 8;
\user begins exploring this layer by hovering over particular vertices to show their labels and immediate neighbors.
\user discovers many interesting quasi-cliques of related words, such as one describing familial relationships (including words like ``daughter,'' ``husband,'' and ``grandparent''), one describing commonly injured body parts (including words like ``knee,'' ``ankle,'' and ``sprained''), and another elongated quasi-clique that describes the levels of negative \textit{surprise} one can experience (including words like ``annoyed,'' ``dismayed,'' and ``mortified'').
\user then clicks on the ``Clone'' toggle, which colors and sizes vertices that are cloned in other layers red (\autoref{fig:scenario}, right).
This reveals two findings: (1), many vertices in the quasi-clique also exist in other layers, showing that these vertices play other roles throughout the graph; and (2), some vertices only exists within this layer, and therefore play a singular role within the entire graph's structure.
\user now has a solid understanding of this graph layer, and wishes to explore another.
Instead of using the \ribbon and clicking on another layer, \user inspects the vertex ``dismayed'' from the negative ``surprise'' quasi-clique from earlier.
\system reveals that it has vertex clones in layers 5 and 3.
\user clicks on the layer 5 clone label for the word ``dismayed,'' and \system adds a visualization of layer 5 underneath the existing layer 8 visualization in the \layers view.
\system focuses on both existences of ``dismayed'' by highlighting its vertex in layer 8 and accompanying clone in layer 5 blue and vertically aligns them in both layers, showing their roles in both graph layers.

\textbf{Local exploration with a global context.}
\user now explores layer 5 starting from ``dismayed.''
By hovering over ``dismayed'' \user notices that its neighbors are similar to the neighbors in layer 8, but are indeed different words (including words like ``angered,'' ``displeased,'' ``embarrassed'').
However, unlike layer 8, ``dismayed'' in layer 5 is connected to a larger connected component, and as \user follows the neighbors of ``dismayed'' throughout the component he notices the words transition from describing one's negative \textit{surprise} to more neutral words, e.g, ``shocked'' and ''surprised.''
Moreover, these neutral \textit{surprise} words form the center of the connected component, and continuing further reveals a new transition from neutral words to positive words such as ``remarkable,'' ``astounding,'' and ``extraordinarily.''
\user has now discovered that words describing \textit{surprise} are represented in this word embedding similar to how humans would think of them: one can be \textit{surprised}, however, the word ``surprised'' itself does not necessarily carry a positive nor negative meaning.
Using \system, we see that neutral words like ``shocked'' and ``surprising'' bridge quasi-cliques of positive and negative \textit{surprise} words together.
While \user performed this exploration by hand, \system automates this by instantly computing approximate shortest paths between selected pairs of vertices in a graph, with the extra ability to then add single vertices to find approximate shortest paths to the already existing path.
We call this representation and mode of exploration the \spnet via sequential egonet expansion, which is later discussed in \autoref{subsec:layers}.
This allows a user to explore semantic information within a connected component of a graph layer, as well as view the information's transition from one side of a connected component to another.

\textbf{Multiple exploration choices.}
Since this component in layer 5 describing one's \textit{surprise} is much larger than the smaller quasi-clique in layer 8, \user now has multiple choices for continuing exploring this word embedding graph using \system:
(1) visit the other connected components in layer 5,
(2) backtrack to layer 8 and use the last clone of ``dismayed'' as a mechanism to perform further cross-layer exploration, or 
(3) return to the beginning and inspect the \overview and \ribbon for a completely different layer to explore.
Regardless of what \user chooses, he can gain a better understanding of the word embedding graph both globally, by visualizing graph layer structure, and locally, by using vertex clones and \spnet representations.

\section{\system: Interactive Large Graph Exploration}
\label{sec:design}

Here we first describe the design challenges motivated by existing work for large graph exploration.
For each challenge, we present our solution that guided the design decision for \system.
The remaining three subsections each describe one of the main coordinated views of \system and highlight their core features for graph sensemaking; these include the 3D \overview (Section \ref{subsec:overview}), the Graph \ribbon (\autoref{subsec:ribbon}), and the \layers view (\autoref{subsec:layers}).

\subsection{Challenges and Design Rationale}
\label{subsec:challeneges}

\begin{itemize}[leftmargin=0mm]
    \setlength\itemsep{1em}
    
    \item[] \textbf{Challenge 1: Variety of overlapping subgraph structure.}
    % \\
    There are a variety of existing techniques that aim to discover structure and patterns in graphs.
    However, while these techniques may find individual structure and patterns, they do not link the findings together, nor do they explain how multiple patterns are associated with one another.
    Revealing such kinds of links between structure and pattern is a hallmark capability that is crucial to sensemaking~\cite{holyoak1997analogical, gentner1997structure}.
    
    \textbf{Our solution:} We utilize the dual nature of graph layers: (1) layers can be explored independently from one another, but more importantly, (2) layers can be linked together using vertex clones as a mechanism of traversal from layer to layer.
    We call this cross-layer exploration (see \autoref{fig:ui}).
        Visualizing the decomposition in 3D may help users more clearly see the overlapping graph structure, which could help them choose which layer of the graph to explore first.
    
    \item[] \textbf{Challenge 2: Local exploration of large graphs.}
    % \\
    Since large graph exploration is difficult from both a visual and computational scalability perspective, querying a graph or considering subgraphs to explore locally can be helpful.
    However, often times the global context is lost using these approaches, as users do not know where in the graph they are exploring, or how different subgraphs are related to one another.
    
    \textbf{Our solution:} We design a novel visual summarization of the edge decomposition called the Graph \ribbon and embed it in the middle of the user interface (\autoref{fig:ui}).
    The \ribbon encodes each layer as a glyph and functions as a global map of the decomposition and graph.
    If a layer is selected to be visualized, a small triangle pointing left or right (denoting if the layer is visualized in the 3D \overview or the \layers view) is displayed next to that layer's glyph.
    We also design novel local exploration techniques within a graph layer (\spnet via sequential egonet expansion) that help users explore graphs locally with a global context. 
    
    \item[] \textbf{Challenge 3: Large graphs.}
    % \\
    While many graphs are small and can be visualized in 2D with standard layouts, many modern graphs are growing increasingly large and complex.
    Not only is this problematic for data visualization itself, but also troublesome for engineering interactive tools.
    The sheer size of the data render many existing visualization tools unusable as they are often designed to visualize the entire graph.

    \textbf{Our solution:} We display a visual summarization of the edge decomposition (called the \ribbon) for a high-level view of the graph and its decomposition.
    Then, we can selectively load and visualize only the layers we desire, skirting scalability challenges that come with visualizing an entire graph at once.
\end{itemize}

\subsection{3D Graph Decomposition Overview}
\label{subsec:overview}

The left view of \system, called the \overview (\autoref{fig:ui}), visualizes graph decompositions in 3D and allows users to zoom, pan, and rotate the 3D structure in-browser and in real-time.
Our edge decomposition divides large graphs into an ordered set of graph layers that is completely dependent upon the topology of the graph.
Since the graph edge set is uniquely partitioned into graph layers, a natural approach to visualize the decomposition is to first perform a traditional 2D layout of the graph in the plane (this assigns vertices $x$ and $y$ coordinates); however, we now assign a $z$ coordinate to each vertex, where the $z$ coordinate is a function of the vertex peel value.
Since graph layers are numerically ordered, when visualizing a decomposition in 3D the highest, most dense layers (e.g., quasi-cliques) rise to the top while the lower layers sink to the bottom (e.g., trees, stars).
\system supports graphs with millions of vertices, but to compute an initial 2D layout is non-trivial; therefore, we use a GPU-accelerated implementation~\cite{brinkmann2017exploiting} of the Barnes-Hutt approximation~\cite{barnes1986hierarchical} to achieve large graph layouts in minutes.

Users can display all graph layers at once or selectively add layers to the \overview.
The \overview also contains options to help users explore and manipulate the 3D structure.
These options include interactive sliders for adjusting the size of the vertices, the height of the layers (e.g., dragging this slider animates splitting the graph into its graph layers), and the spread of the layers (i.e., scaling the $x$ and $y$ positions of the nodes).
The ``Animate'' button simply automates dragging the height slider to watch a short animation of the original 2D graph dividing into its 3D decomposition.
Since navigating large 3D structures suffers from a distorted perspective, a ``Top View'' button is present to return the camera to its original position.
This 3D \overview naturally visualizes how graphs decompose into layers and highlights how vertices can be cloned throughout multiple layers; if a vertex has clones, they will be stacked vertically along the $z$-axis (see the two blue vertex clones for ``caeciliidae'' in \autoref{fig:ui}, left).
Lastly, in the right view of \system, discussed in \autoref{subsec:layers}, vertices can be selected to perform various tasks.
When a vertex is selected, it is highlighted blue, as seen in \autoref{fig:ui} on the right where the vertex ``caeciliidae' is selected.
We link the state from the \overview and the \layers view of \system, i.e., every selected vertex in the \layers view is also highlighted blue in the \overview (see the call out in \autoref{fig:ui}, left) , so users can always refer back to the 3D structure to see which vertices they have selected.

\subsection{Graph Ribbon: Edge Decomposition Summarization}
\label{subsec:ribbon}

For each layer produced by the edge decomposition, we compute a set of measures that together provide a quantitative summary of the edge decomposition.
We encode these measures for every layer as a horizontal bar glyph to create the visualization in the middle view of the \system user interface, called the \ribbon (\autoref{fig:ui}).
While there are many diverse graph measures originating from graph theory, graph mining, and network science, we pick five distinct measures we think summarize the graph well; however, it should be noted that other measures can be computed for each layer and included in the \ribbon visualization for further analysis or specialized tasks.
While inspecting graph measures on each layer independently can be enlightening, visualizing each metric across layers as a distribution highlights the power of the edge decomposition.
Hovering over a layer displays a tooltip with the five computed measures displayed as numerical values for a given layer.
The top of \ribbon includes a menu button that contains options to toggle each of the visualized measures, as well as a linear / log scale toggle for the axis.

The \ribbon is not only a summarization of the edge decomposition; clicking on a specific layer's glyph displays that layer on the right of the user interface, discussed in detail in \autoref{subsec:layers}, while a Command+Click displays that layer in 3D in the \overview.
Lastly, the entire \ribbon can be dragged using either of the arrows at the top to give more screen real estate to either the \overview or \layers view.
Listed below are the five measures and how they are visualized in \system.

\begin{figure}[tb]
 \centering
 \includegraphics[width=1.0\columnwidth]{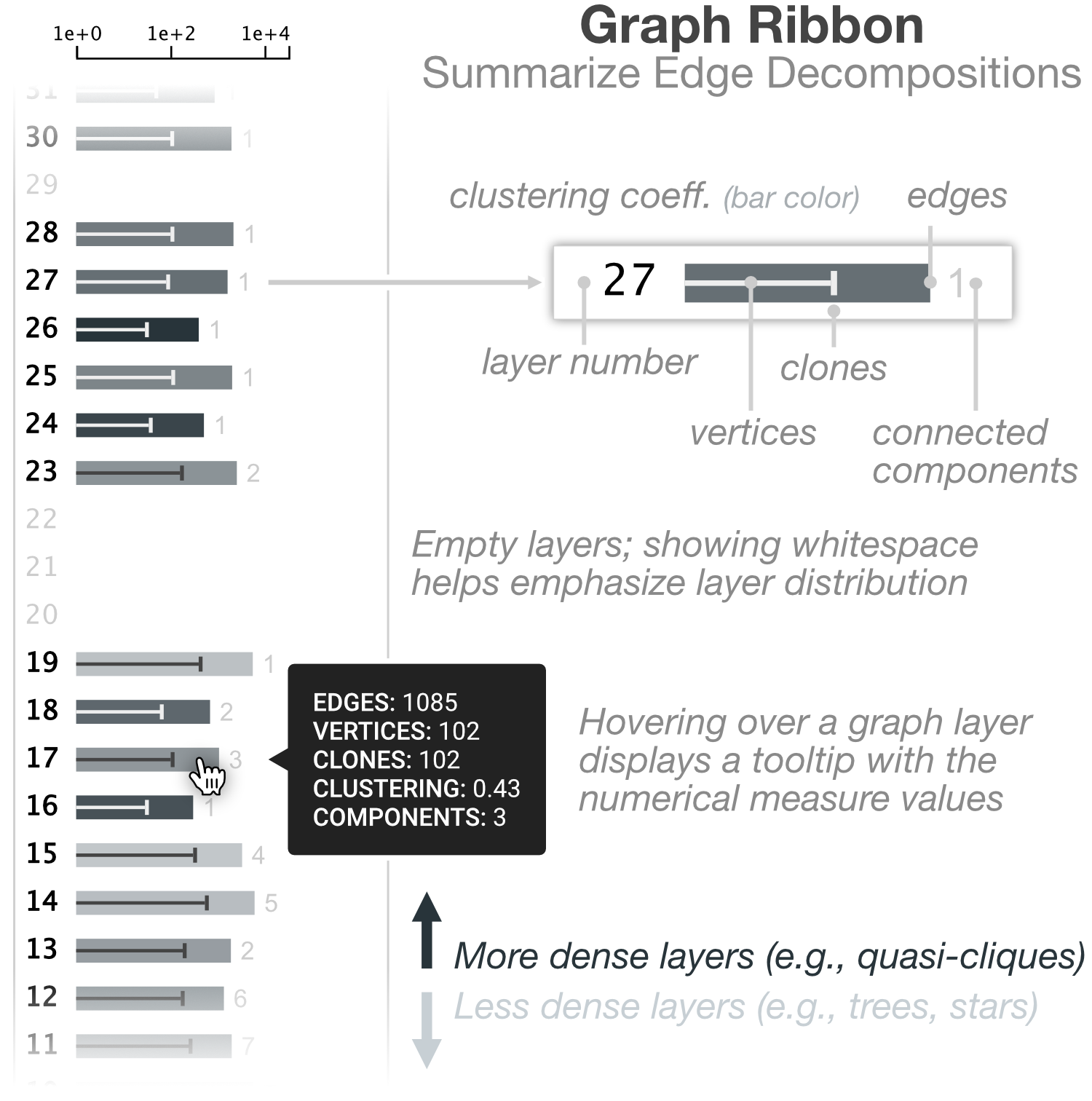}
 \caption{
 The \ribbon for the Wikipedia GloVe word embedding graph.
 The graph \ribbon summarizes the edge decomposition using graph measures such as the vertex count, the edge count, the cloned vertex count, clustering coefficient, and number of connected components.
 }
 \label{fig:ribbon}
\end{figure}

\subsection{Navigating and Exploring Graphs Using Graph Layers and Vertex Clones}
\label{subsec:layers}
The last of the three main views of \system is the \layers view (\autoref{fig:ui}, right).
When a layer in the \ribbon is clicked, \system visualizes that specific layer as an interactive node-link diagram.
This visualization is completely interactive: users can zoom and pan on the graph, as well as drag, pin, and select specific vertices.
Hovering over a vertex highlights it, its edges, and its neighbors orange (\autoref{fig:ui}, right).
The computed layer measures are listed in the top left corner of the \layers view.
If the specified layer only contains a single connected component, a message is shown displaying how many edges the component requires to become a complete clique.
Conversely, if the specified layer contains multiple connected components, a different message is shown displaying the largest connected component's vertex and edge count; a slider is also shown that hides components in order of their size, i.e., dragging the slider from left to right hides the smallest connected components, eventually showing the only the largest component in the graph layer.

\textbf{Independent graph layer layouts.}
\system supports multiple interactions for exploring within a single layer.
Toggles are present for showing and hiding the vertices and edges of the layer.
The ``Redraw'' toggle animates the layer unraveling using a precomputed independent force-directed layout to better show the decomposition's found structure (\autoref{fig:scenario}).
However, users can also run a force-directed layout in-browser by clicking the ``Live Layout'' toggle; the layout computation continues until the toggle is turned off.
This can be useful for computing a larger connected component's layout within a layer; by using the component slider to hide smaller components the desired larger component can be redrawn independently for better structural clarity.

\textbf{Graph layer contour motifs.}
The ``Motif'' toggle, when turned on, computes a contour map of the graph layer by performing kernel-density estimation (KDE) on the vertices of the layer.
Options for adjusting the bandwidth and number of thresholds for the KDE are present underneath the toggle.
This contour motif provides a higher-level, more abstract representation of a graph layer, creating a proxy for vertex density~\cite{lin2010contextour, cao2010facetatlas}.
The contour motif is also instantly recomputed whenever a user drags a vertex or uses one of the above interactions to re-redraw a layer.

{\renewcommand{\arraystretch}{1.1}
\begin{table*}[tb]
  \caption{
  Results for our fast and scalable edge decomposition algorithm across of number of different graphs varying in size and domain.
  Experimental timings are the average of 5 runs for each graph.
  We can decompose graphs with millions of edges in seconds, and graphs with hundreds of millions of edges in minutes.
  }
  \label{tab:algorithm-results}
	\centering
  \begin{tabular}{llrrrrr}
  \toprule
  \textbf{Graph} & \textbf{Graph Type} & \textbf{Vertices} & \textbf{Edges} & \textbf{Time} (sec.) & \textbf{Layers} & \textbf{Highest Peel} 
  \\
  \midrule
  Bible Names             & co-occurrence  & 1,774   & 9,131       & 0.01   & 12 & 15  \\
  Google+                 & social network & 23,628  & 39,242      &   0.02     & 10  & 13  \\
  arXiv astro-ph                 & co-authorship  & 18,771  & 198,050     & 0.10   & 47 & 56  \\
  Amazon                  & co-purchase & 334,863    & 925,872     &   0.12     & 6 & 6 \\
  US Patents              & citation network & 3,774,768  & 16,518,947  & 11.73  & 41 & 64 \\
  Pokec                & social network & 1,632,803  & 30,622,564  & 12.33  & 44 & 70  \\
  LiveJournal            & social network & 4,847,571  & 68,993,773  & 120.70 & 179 & 510  \\
  Wikipedia Links (German) & hyperlink network & 3,225,565  & 81,626,917  & 225.40 & 320 & 1656 \\
  Orkut                    & social network    & 3,072,441  & 117,184,899 & 91.84 & 91 & 253 \\
  \bottomrule
  \end{tabular}
\end{table*}
}

\textbf{Shortest-path-nets via sequential egonet expansion.}
The ``Path'' button allows users to explore a single graph layer by building a \spnet representation.
When two vertices are selected within a graph layer, clicking the ``Path'' button will compute the shortest path between the vertices, or an approximation depending on the component size, highlight the computed path blue, and display the vertex labels along this path.
A user can now select a third vertex somewhere else in the layer and click the ``Path'' once again to find an approximate shortest path from the third vertex to any other vertex along the existing path; iterating this process computes what we call a \spnet via sequential egonet expansion.
This mode of exploration is especially useful for observing the transition of semantic information from one side of a large connected component to another.

\textbf{Vertex clones.}
Lastly, the ``Clone'' toggle shows which vertices of a graph layer are clones or not.
When toggled on, \system colors cloned vertices red and sizes each vertex according to how many clones that vertex has in the entire graph (see \autoref{fig:scenario}).
When locally exploring a single graph layer, visualizing the vertex clones provides global context for how a particular vertex may participate in many graph layers at once.
Conversely, vertices that do not have any clones remain colored gray, and stand out as ``secret agents'' within a particular layer.
These vertices are equally informative, as all of their edges exist within a single layer, indicating that they play a singular role in the graph.
Hovering over a vertex displays its label and lists the other layers its clones exist in.
If a user clicks on one of the clones in the list, \system shows the selected layer underneath the original visualized layer and centers each of their displays on the selected vertex and its clone (see \autoref{fig:ui}, right).
These vertices are now selected and synced, i.e., dragging one of the vertices will also drag the other, updating their position in both layers, reinforcing the notion that a single vertex can participate and influence multiple layers throughout an entire graph.
For example, in \autoref{fig:ui} on the right, we see the blue vertex ``caeciliidae'' (a worm-like amphibian) in layer 30 bridges two quasi-cliques (families of birds and families of sea snails) together, while its clone in layer 25 participates in another single quasi-clique (families of land creatures).

\section{Algorithm Results and System Design}
\label{sec:technical}

Recall our edge decomposition simultaneously reveals (1) peculiar subgraph structure discovered through the decomposition's layers, (e.g., quasi-cliques, multi-partite-cores), and (2) possible vertex roles in linking such subgraph patterns across layers.
We utilize the edge decomposition based on fixed points of degree peeling by Abello et al.~\cite{abello2014network} and make improvements to increase its performance, both in computation speed and scalability.

\subsection{Large Graph Decomposition Experimental Results}
Our fast and scalable edge decomposition is implemented in C++; however, we improve performance by leveraging memory mapping~\cite{lin2014mmap} to load large graphs into memory.
Recall that the edge decomposition runs traditional k-core decomposition $L$ times, where $L$ is the number of layers in the graph; therefore, we use a recent multithreaded implementation of k-core decomposition to achieve significant speedup~\cite{kabir2017parallel}.

We report results on decomposing graphs using our fast and scalable implementation.
We chose a wide range of of graphs, varying in both size (e.g. thousands to hundreds of millions of edges) and domain (e.g. social networks, hyperlink networks, and co-occurrence networks).
We performed our experiments on a single commodity computer equipped with an Intel i7 6-core processor clocked at 3.3GHz and 32GB of RAM.
For each graph, the timing result is averaged over 5 runs.
All results are tabulated in \autoref{tab:algorithm-results}, which includes the graph, its vertex and edge count, the algorithm compute time without preprocessing steps, the number of layers each graph produces, and the highest peel value from the decomposition (since a graph with $L$ layers does not necessarily mean the $L$ layers correspond to $[1,2,3, \dots, L]$).
We can decompose graphs with millions of edges in seconds, and graphs with hundreds of millions of edges in minutes.

\subsection{System Design}
For graph drawing, we use a force-directed technique to layout the original graph; however, in order to calculate the $(x,y)$ coordinates of every vertex in large graphs where typical force-directed layouts are slow and expensive to compute, we perform the layout computation using the Barnes-Hutt~\cite{barnes1986hierarchical} approximation on a GPU for significant speedup~\cite{brinkmann2017exploiting}.
Computing the edge decomposition of our graph and the global graph layout are independent computations.
When both are completed, we process their output together using Python to compute graph layer measures, vertex clones, and format the data to be ingested by \system.
The visualization system is web-based and uses the latest JavaScript libraries to render elements to the screen, such as the now ubiquitously used D3~\cite{bostock2011d3} for manipulating SVGs and the GPU-powered library three.js (\url{https://threejs.org/}) for rendering the 3D graphics.

\begin{acks}
This work was supported in part by NSF grants IIS-1563816, IIS-1363971, IIS-1217559, TWC-1526254, and CNS 17041701.
This work was supported by a NASA Space Technology Research Fellowship.
\end{acks}

\bibliographystyle{ACM-Reference-Format}
\bibliography{main}

%%% -*-BibTeX-*-
%%% Do NOT edit. File created by BibTeX with style
%%% ACM-Reference-Format-Journals [18-Jan-2012].

\begin{thebibliography}{28}

%%% ====================================================================
%%% NOTE TO THE USER: you can override these defaults by providing
%%% customized versions of any of these macros before the \bibliography
%%% command.  Each of them MUST provide its own final punctuation,
%%% except for \shownote{}, \showDOI{}, and \showURL{}.  The latter two
%%% do not use final punctuation, in order to avoid confusing it with
%%% the Web address.
%%%
%%% To suppress output of a particular field, define its macro to expand
%%% to an empty string, or better, \unskip, like this:
%%%
%%% \newcommand{\showDOI}[1]{\unskip}   % LaTeX syntax
%%%
%%% \def \showDOI #1{\unskip}           % plain TeX syntax
%%%
%%% ====================================================================

\ifx \showCODEN    \undefined \def \showCODEN     #1{\unskip}     \fi
\ifx \showDOI      \undefined \def \showDOI       #1{#1}\fi
\ifx \showISBNx    \undefined \def \showISBNx     #1{\unskip}     \fi
\ifx \showISBNxiii \undefined \def \showISBNxiii  #1{\unskip}     \fi
\ifx \showISSN     \undefined \def \showISSN      #1{\unskip}     \fi
\ifx \showLCCN     \undefined \def \showLCCN      #1{\unskip}     \fi
\ifx \shownote     \undefined \def \shownote      #1{#1}          \fi
\ifx \showarticletitle \undefined \def \showarticletitle #1{#1}   \fi
\ifx \showURL      \undefined \def \showURL       {\relax}        \fi
% The following commands are used for tagged output and should be
% invisible to TeX
\providecommand\bibfield[2]{#2}
\providecommand\bibinfo[2]{#2}
\providecommand\natexlab[1]{#1}
\providecommand\showeprint[2][]{arXiv:#2}

\bibitem[\protect\citeauthoryear{Abello, Hohman, and Chau}{Abello
  et~al\mbox{.}}{2015}]%
        {hohman2017playground}
\bibfield{author}{\bibinfo{person}{James Abello}, \bibinfo{person}{Fred
  Hohman}, {and} \bibinfo{person}{Duen~Horng Chau}.}
  \bibinfo{year}{2015}\natexlab{}.
\newblock \showarticletitle{3D Exploration of Graph Layers via Vertex Cloning}.
  In \bibinfo{booktitle}{\emph{2017 IEEE Conference on Visual Analytics Science
  and Technology (VAST), Poster}}. IEEE.
\newblock


\bibitem[\protect\citeauthoryear{Abello and Queyroi}{Abello and
  Queyroi}{2014}]%
        {abello2014network}
\bibfield{author}{\bibinfo{person}{James Abello} {and}
  \bibinfo{person}{Fran{\c{c}}ois Queyroi}.} \bibinfo{year}{2014}\natexlab{}.
\newblock \showarticletitle{Network decomposition into fixed points of degree
  peeling}.
\newblock \bibinfo{journal}{\emph{Social Network Analysis and Mining}}
  \bibinfo{volume}{4}, \bibinfo{number}{1} (\bibinfo{year}{2014}),
  \bibinfo{pages}{1--14}.
\newblock


\bibitem[\protect\citeauthoryear{Abello, Van~Ham, and Krishnan}{Abello
  et~al\mbox{.}}{2006}]%
        {abello2006ask}
\bibfield{author}{\bibinfo{person}{James Abello}, \bibinfo{person}{Frank
  Van~Ham}, {and} \bibinfo{person}{Neeraj Krishnan}.}
  \bibinfo{year}{2006}\natexlab{}.
\newblock \showarticletitle{Ask-graphview: A large scale graph visualization
  system}.
\newblock \bibinfo{journal}{\emph{IEEE transactions on visualization and
  computer graphics}} \bibinfo{volume}{12}, \bibinfo{number}{5}
  (\bibinfo{year}{2006}), \bibinfo{pages}{669--676}.
\newblock


\bibitem[\protect\citeauthoryear{Alper, Bach, Henry~Riche, Isenberg, and
  Fekete}{Alper et~al\mbox{.}}{2013}]%
        {alper2013weighted}
\bibfield{author}{\bibinfo{person}{Basak Alper}, \bibinfo{person}{Benjamin
  Bach}, \bibinfo{person}{Nathalie Henry~Riche}, \bibinfo{person}{Tobias
  Isenberg}, {and} \bibinfo{person}{Jean-Daniel Fekete}.}
  \bibinfo{year}{2013}\natexlab{}.
\newblock \showarticletitle{Weighted graph comparison techniques for brain
  connectivity analysis}. In \bibinfo{booktitle}{\emph{Proceedings of the
  SIGCHI Conference on Human Factors in Computing Systems}}. ACM,
  \bibinfo{pages}{483--492}.
\newblock


\bibitem[\protect\citeauthoryear{Antol, Agrawal, Lu, Mitchell, Batra,
  Lawrence~Zitnick, and Parikh}{Antol et~al\mbox{.}}{2015}]%
        {antol2015vqa}
\bibfield{author}{\bibinfo{person}{Stanislaw Antol}, \bibinfo{person}{Aishwarya
  Agrawal}, \bibinfo{person}{Jiasen Lu}, \bibinfo{person}{Margaret Mitchell},
  \bibinfo{person}{Dhruv Batra}, \bibinfo{person}{C Lawrence~Zitnick}, {and}
  \bibinfo{person}{Devi Parikh}.} \bibinfo{year}{2015}\natexlab{}.
\newblock \showarticletitle{Vqa: Visual question answering}. In
  \bibinfo{booktitle}{\emph{Proceedings of the IEEE International Conference on
  Computer Vision}}. \bibinfo{pages}{2425--2433}.
\newblock


\bibitem[\protect\citeauthoryear{Archambault, Munzner, and Auber}{Archambault
  et~al\mbox{.}}{2007a}]%
        {archambault2007grouse}
\bibfield{author}{\bibinfo{person}{Daniel Archambault}, \bibinfo{person}{Tamara
  Munzner}, {and} \bibinfo{person}{David Auber}.}
  \bibinfo{year}{2007}\natexlab{a}.
\newblock \showarticletitle{Grouse: Feature-Based, Steerable Graph Hierarchy
  Exploration.}. In \bibinfo{booktitle}{\emph{EuroVis}},
  Vol.~\bibinfo{volume}{2007}. \bibinfo{pages}{67--74}.
\newblock


\bibitem[\protect\citeauthoryear{Archambault, Munzner, and Auber}{Archambault
  et~al\mbox{.}}{2007b}]%
        {archambault2007topolayout}
\bibfield{author}{\bibinfo{person}{Daniel Archambault}, \bibinfo{person}{Tamara
  Munzner}, {and} \bibinfo{person}{David Auber}.}
  \bibinfo{year}{2007}\natexlab{b}.
\newblock \showarticletitle{Topolayout: Multilevel graph layout by topological
  features}.
\newblock \bibinfo{journal}{\emph{IEEE transactions on visualization and
  computer graphics}} \bibinfo{volume}{13}, \bibinfo{number}{2}
  (\bibinfo{year}{2007}).
\newblock


\bibitem[\protect\citeauthoryear{Bahdanau, Cho, and Bengio}{Bahdanau
  et~al\mbox{.}}{2015}]%
        {bahdanau2014neural}
\bibfield{author}{\bibinfo{person}{Dzmitry Bahdanau},
  \bibinfo{person}{Kyunghyun Cho}, {and} \bibinfo{person}{Yoshua Bengio}.}
  \bibinfo{year}{2015}\natexlab{}.
\newblock \showarticletitle{Neural machine translation by jointly learning to
  align and translate}.
\newblock \bibinfo{journal}{\emph{International Conference on Learning
  Representations (ICLR)}} (\bibinfo{year}{2015}).
\newblock


\bibitem[\protect\citeauthoryear{Barnes and Hut}{Barnes and Hut}{1986}]%
        {barnes1986hierarchical}
\bibfield{author}{\bibinfo{person}{Josh Barnes} {and} \bibinfo{person}{Piet
  Hut}.} \bibinfo{year}{1986}\natexlab{}.
\newblock \showarticletitle{A hierarchical O (N log N) force-calculation
  algorithm}.
\newblock \bibinfo{journal}{\emph{nature}} \bibinfo{volume}{324},
  \bibinfo{number}{6096} (\bibinfo{year}{1986}), \bibinfo{pages}{446}.
\newblock


\bibitem[\protect\citeauthoryear{Bengio, Ducharme, Vincent, and Jauvin}{Bengio
  et~al\mbox{.}}{2003}]%
        {bengio2003neural}
\bibfield{author}{\bibinfo{person}{Yoshua Bengio}, \bibinfo{person}{R{\'e}jean
  Ducharme}, \bibinfo{person}{Pascal Vincent}, {and} \bibinfo{person}{Christian
  Jauvin}.} \bibinfo{year}{2003}\natexlab{}.
\newblock \showarticletitle{A neural probabilistic language model}.
\newblock \bibinfo{journal}{\emph{Journal of machine learning research}}
  \bibinfo{volume}{3}, \bibinfo{number}{Feb} (\bibinfo{year}{2003}),
  \bibinfo{pages}{1137--1155}.
\newblock


\bibitem[\protect\citeauthoryear{Bostock, Ogievetsky, and Heer}{Bostock
  et~al\mbox{.}}{2011}]%
        {bostock2011d3}
\bibfield{author}{\bibinfo{person}{Michael Bostock}, \bibinfo{person}{Vadim
  Ogievetsky}, {and} \bibinfo{person}{Jeffrey Heer}.}
  \bibinfo{year}{2011}\natexlab{}.
\newblock \showarticletitle{D$^3$ data-driven documents}.
\newblock \bibinfo{journal}{\emph{IEEE transactions on visualization and
  computer graphics}} \bibinfo{volume}{17}, \bibinfo{number}{12}
  (\bibinfo{year}{2011}), \bibinfo{pages}{2301--2309}.
\newblock


\bibitem[\protect\citeauthoryear{Brinkmann, Rietveld, and Takes}{Brinkmann
  et~al\mbox{.}}{2017}]%
        {brinkmann2017exploiting}
\bibfield{author}{\bibinfo{person}{Govert~G Brinkmann},
  \bibinfo{person}{Kristian~FD Rietveld}, {and} \bibinfo{person}{Frank~W
  Takes}.} \bibinfo{year}{2017}\natexlab{}.
\newblock \showarticletitle{Exploiting GPUs for fast force-directed
  visualization of large-scale networks}. In \bibinfo{booktitle}{\emph{Parallel
  Processing (ICPP), 2017 46th International Conference on}}. IEEE,
  \bibinfo{pages}{382--391}.
\newblock


\bibitem[\protect\citeauthoryear{Cao, Sun, Lin, Gotz, Liu, and Qu}{Cao
  et~al\mbox{.}}{2010}]%
        {cao2010facetatlas}
\bibfield{author}{\bibinfo{person}{Nan Cao}, \bibinfo{person}{Jimeng Sun},
  \bibinfo{person}{Yu-Ru Lin}, \bibinfo{person}{David Gotz},
  \bibinfo{person}{Shixia Liu}, {and} \bibinfo{person}{Huamin Qu}.}
  \bibinfo{year}{2010}\natexlab{}.
\newblock \showarticletitle{Facetatlas: Multifaceted visualization for rich
  text corpora}.
\newblock \bibinfo{journal}{\emph{IEEE transactions on visualization and
  computer graphics}} \bibinfo{volume}{16}, \bibinfo{number}{6}
  (\bibinfo{year}{2010}), \bibinfo{pages}{1172--1181}.
\newblock


\bibitem[\protect\citeauthoryear{Chau, Kittur, Hong, and Faloutsos}{Chau
  et~al\mbox{.}}{2011}]%
        {chau2011apolo}
\bibfield{author}{\bibinfo{person}{Duen~Horng Chau}, \bibinfo{person}{Aniket
  Kittur}, \bibinfo{person}{Jason~I Hong}, {and} \bibinfo{person}{Christos
  Faloutsos}.} \bibinfo{year}{2011}\natexlab{}.
\newblock \showarticletitle{Apolo: making sense of large network data by
  combining rich user interaction and machine learning}. In
  \bibinfo{booktitle}{\emph{Proceedings of the SIGCHI Conference on Human
  Factors in Computing Systems}}. ACM, \bibinfo{pages}{167--176}.
\newblock


\bibitem[\protect\citeauthoryear{Cui, Zhou, Qu, Wong, and Li}{Cui
  et~al\mbox{.}}{2008}]%
        {cui2008geometry}
\bibfield{author}{\bibinfo{person}{Weiwei Cui}, \bibinfo{person}{Hong Zhou},
  \bibinfo{person}{Huamin Qu}, \bibinfo{person}{Pak~Chung Wong}, {and}
  \bibinfo{person}{Xiaoming Li}.} \bibinfo{year}{2008}\natexlab{}.
\newblock \showarticletitle{Geometry-based edge clustering for graph
  visualization}.
\newblock \bibinfo{journal}{\emph{IEEE Transactions on Visualization and
  Computer Graphics}} \bibinfo{volume}{14}, \bibinfo{number}{6}
  (\bibinfo{year}{2008}), \bibinfo{pages}{1277--1284}.
\newblock


\bibitem[\protect\citeauthoryear{Dunne and Shneiderman}{Dunne and
  Shneiderman}{2013}]%
        {dunne2013motif}
\bibfield{author}{\bibinfo{person}{Cody Dunne} {and} \bibinfo{person}{Ben
  Shneiderman}.} \bibinfo{year}{2013}\natexlab{}.
\newblock \showarticletitle{Motif simplification: improving network
  visualization readability with fan, connector, and clique glyphs}. In
  \bibinfo{booktitle}{\emph{CHI}}. ACM, \bibinfo{pages}{3247--3256}.
\newblock


\bibitem[\protect\citeauthoryear{Gentner and Markman}{Gentner and
  Markman}{1997}]%
        {gentner1997structure}
\bibfield{author}{\bibinfo{person}{Dedre Gentner} {and}
  \bibinfo{person}{Arthur~B Markman}.} \bibinfo{year}{1997}\natexlab{}.
\newblock \showarticletitle{Structure mapping in analogy and similarity.}
\newblock \bibinfo{journal}{\emph{American psychologist}} \bibinfo{volume}{52},
  \bibinfo{number}{1} (\bibinfo{year}{1997}), \bibinfo{pages}{45}.
\newblock


\bibitem[\protect\citeauthoryear{Heer and Card}{Heer and Card}{2004}]%
        {heer2004doitrees}
\bibfield{author}{\bibinfo{person}{Jeffrey Heer} {and}
  \bibinfo{person}{Stuart~K Card}.} \bibinfo{year}{2004}\natexlab{}.
\newblock \showarticletitle{DOITrees revisited: scalable, space-constrained
  visualization of hierarchical data}. In \bibinfo{booktitle}{\emph{Proceedings
  of the working conference on Advanced visual interfaces}}. ACM,
  \bibinfo{pages}{421--424}.
\newblock


\bibitem[\protect\citeauthoryear{Holten}{Holten}{2006}]%
        {holten2006hierarchical}
\bibfield{author}{\bibinfo{person}{Danny Holten}.}
  \bibinfo{year}{2006}\natexlab{}.
\newblock \showarticletitle{Hierarchical edge bundles: Visualization of
  adjacency relations in hierarchical data}.
\newblock \bibinfo{journal}{\emph{IEEE Transactions on visualization and
  computer graphics}} \bibinfo{volume}{12}, \bibinfo{number}{5}
  (\bibinfo{year}{2006}), \bibinfo{pages}{741--748}.
\newblock


\bibitem[\protect\citeauthoryear{Holyoak and Thagard}{Holyoak and
  Thagard}{1997}]%
        {holyoak1997analogical}
\bibfield{author}{\bibinfo{person}{Keith~J Holyoak} {and} \bibinfo{person}{Paul
  Thagard}.} \bibinfo{year}{1997}\natexlab{}.
\newblock \showarticletitle{The analogical mind.}
\newblock \bibinfo{journal}{\emph{American psychologist}} \bibinfo{volume}{52},
  \bibinfo{number}{1} (\bibinfo{year}{1997}), \bibinfo{pages}{35}.
\newblock


\bibitem[\protect\citeauthoryear{Kabir and Madduri}{Kabir and Madduri}{2017}]%
        {kabir2017parallel}
\bibfield{author}{\bibinfo{person}{Humayun Kabir} {and} \bibinfo{person}{Kamesh
  Madduri}.} \bibinfo{year}{2017}\natexlab{}.
\newblock \showarticletitle{Parallel k-core decomposition on multicore
  platforms}. In \bibinfo{booktitle}{\emph{Parallel and Distributed Processing
  Symposium Workshops (IPDPSW), 2017 IEEE International}}. IEEE,
  \bibinfo{pages}{1482--1491}.
\newblock


\bibitem[\protect\citeauthoryear{Koutra, Kang, Vreeken, and Faloutsos}{Koutra
  et~al\mbox{.}}{2014}]%
        {koutra2014vog}
\bibfield{author}{\bibinfo{person}{Danai Koutra}, \bibinfo{person}{U Kang},
  \bibinfo{person}{Jilles Vreeken}, {and} \bibinfo{person}{Christos
  Faloutsos}.} \bibinfo{year}{2014}\natexlab{}.
\newblock \showarticletitle{VOG: summarizing and understanding large graphs}.
  In \bibinfo{booktitle}{\emph{SDM}}. SIAM, \bibinfo{pages}{91--99}.
\newblock


\bibitem[\protect\citeauthoryear{Lin, Sun, Cao, and Liu}{Lin
  et~al\mbox{.}}{2010}]%
        {lin2010contextour}
\bibfield{author}{\bibinfo{person}{Yu-Ru Lin}, \bibinfo{person}{Jimeng Sun},
  \bibinfo{person}{Nan Cao}, {and} \bibinfo{person}{Shixia Liu}.}
  \bibinfo{year}{2010}\natexlab{}.
\newblock \showarticletitle{Contextour: Contextual contour visual analysis on
  dynamic multi-relational clustering}. In
  \bibinfo{booktitle}{\emph{Proceedings of the 2010 SIAM International
  Conference on Data Mining}}. SIAM, \bibinfo{pages}{418--429}.
\newblock


\bibitem[\protect\citeauthoryear{Lin, Kahng, Sabrin, Chau, Lee, and Kang}{Lin
  et~al\mbox{.}}{2014}]%
        {lin2014mmap}
\bibfield{author}{\bibinfo{person}{Zhiyuan Lin}, \bibinfo{person}{Minsuk
  Kahng}, \bibinfo{person}{Kaeser~Md Sabrin}, \bibinfo{person}{Duen Horng~Polo
  Chau}, \bibinfo{person}{Ho Lee}, {and} \bibinfo{person}{U Kang}.}
  \bibinfo{year}{2014}\natexlab{}.
\newblock \showarticletitle{Mmap: Fast billion-scale graph computation on a pc
  via memory mapping}. In \bibinfo{booktitle}{\emph{Big Data (Big Data), 2014
  IEEE International Conference on}}. IEEE, \bibinfo{pages}{159--164}.
\newblock


\bibitem[\protect\citeauthoryear{Mikolov, Chen, Corrado, and Dean}{Mikolov
  et~al\mbox{.}}{2013}]%
        {mikolov2013efficient}
\bibfield{author}{\bibinfo{person}{Tomas Mikolov}, \bibinfo{person}{Kai Chen},
  \bibinfo{person}{Greg Corrado}, {and} \bibinfo{person}{Jeffrey Dean}.}
  \bibinfo{year}{2013}\natexlab{}.
\newblock \showarticletitle{Efficient estimation of word representations in
  vector space}.
\newblock \bibinfo{journal}{\emph{arXiv preprint arXiv:1301.3781}}
  (\bibinfo{year}{2013}).
\newblock


\bibitem[\protect\citeauthoryear{Pennington, Socher, and Manning}{Pennington
  et~al\mbox{.}}{2014}]%
        {pennington2014glove}
\bibfield{author}{\bibinfo{person}{Jeffrey Pennington},
  \bibinfo{person}{Richard Socher}, {and} \bibinfo{person}{Christopher
  Manning}.} \bibinfo{year}{2014}\natexlab{}.
\newblock \showarticletitle{Glove: Global vectors for word representation}. In
  \bibinfo{booktitle}{\emph{Proceedings of the 2014 conference on empirical
  methods in natural language processing (EMNLP)}}.
  \bibinfo{pages}{1532--1543}.
\newblock


\bibitem[\protect\citeauthoryear{Pienta, Kahng, Lin, Vreeken, Talukdar, Abello,
  Parameswaran, and Chau}{Pienta et~al\mbox{.}}{2017}]%
        {pienta2017facets}
\bibfield{author}{\bibinfo{person}{Robert Pienta}, \bibinfo{person}{Minsuk
  Kahng}, \bibinfo{person}{Zhiyuan Lin}, \bibinfo{person}{Jilles Vreeken},
  \bibinfo{person}{Partha Talukdar}, \bibinfo{person}{James Abello},
  \bibinfo{person}{Ganesh Parameswaran}, {and} \bibinfo{person}{Duen~Horng
  Chau}.} \bibinfo{year}{2017}\natexlab{}.
\newblock \showarticletitle{Facets: Adaptive local exploration of large
  graphs}. In \bibinfo{booktitle}{\emph{Proceedings of the 2017 SIAM
  International Conference on Data Mining}}. SIAM, \bibinfo{pages}{597--605}.
\newblock


\bibitem[\protect\citeauthoryear{Sahu, Mhedhbi, Salihoglu, Lin, and
  {\"O}zsu}{Sahu et~al\mbox{.}}{2017}]%
        {sahu2017ubiquity}
\bibfield{author}{\bibinfo{person}{Siddhartha Sahu}, \bibinfo{person}{Amine
  Mhedhbi}, \bibinfo{person}{Semih Salihoglu}, \bibinfo{person}{Jimmy Lin},
  {and} \bibinfo{person}{M~Tamer {\"O}zsu}.} \bibinfo{year}{2017}\natexlab{}.
\newblock \showarticletitle{The Ubiquity of Large Graphs and Surprising
  Challenges of Graph Processing}.
\newblock \bibinfo{journal}{\emph{Proceedings of the VLDB Endowment}}
  \bibinfo{volume}{11}, \bibinfo{number}{4} (\bibinfo{year}{2017}).
\newblock


\end{thebibliography}

\end{document}